\documentclass[twocolumn,showpacs,amsmath,amssymb]{revtex4}
    \usepackage{graphicx}
    \usepackage{color}
    \usepackage{dcolumn}
    \usepackage{bm}
    \usepackage{xspace}

    \newcommand{\schr}{Schr\"odinger\xspace}

    \newcommand{\IgnoreThis}[1]{#1}%_this_does_NOT_ignore_text

\begin{document}

\title{Equivalence between free and harmonically trapped quantum particles}

\author{Ole Steuernagel}%$^\dag$}
%aps%\affiliation{School of Physics, Astronomy and Mathematics,
%aps%University of Hertfordshire, Hatfield, AL10 9AB, UK }

\address{School of Physics, Astronomy and Mathematics, University of
Hertfordshire, Hatfield, AL10 9AB, UK }
\email{O.Steuernagel@herts.ac.uk}%aps%

\date{\today}

\begin{abstract}
  It is shown that general solutions of the free-particle \schr
  equation can be mapped onto solutions of the \schr equation for the
  harmonic oscillator. This is done in such a way that the time
  evolution of a free particle subjected to a sudden
  transition to a harmonic potential can be described by a simple
  coordinate transformation applied at the transition time. This
  procedure is computationally more efficient than either
  state-projection or propagator techniques. A concatenation
  of the map and its inverse allows us to map from one harmonic
  oscillator to another with a different spring constant.
\end{abstract}

\pacs{
03.65.Ca, %Formalism
03.65.Db, %Functional analytical methods
03.65.Fd, %Algebraic methods (see also 02.20.-a Group theory)
03.65.Ge  %Solutions of wave equations: bound states
}
%aps%

\maketitle

\IgnoreThis{\section{Introduction}}

It is known that laser beam modes in the paraxial approximation are
equivalent to the eigenmodes of the harmonic
oscillator~\cite{Nienhuis_PRA93,Ole_AJP05}. The corresponding
(one-dimensional, scalar) wave equation
\begin{eqnarray}
{\frac {\partial^2 }{\partial  x^2  }}
 \theta \left( x,z \right)  +i
\alpha\,{\frac {\partial }{\partial z}}  \theta \left( x,z
 \right)
 =0     \label{eq_ParaxWaveEq}
\end{eqnarray}
arises as an approximation to Maxwell's
equations~\cite{Haus.book_00}. In this optical context the parameter
$\alpha = 4 \pi / \lambda$ is proportional to the (stationary,
monochromatic beam's) wave number or inversely proportional to the
light's wavelength~$\lambda$.

The \schr equation for a free particle in one spatial dimension~$x$ is
given by
\begin{eqnarray}
{\frac {\partial^2 }{\partial  x^2  }}
 \phi \left( x,t \right)  +i
\alpha\,{\frac {\partial }{\partial t}}  \phi \left( x,t
 \right)
 =0   \label{eq_FreeSchroedingerEq}
\end{eqnarray}
where the parameter $\alpha = 2 M / \hbar$ contains the particle mass
$M$ and Planck's constant $h$, namely $\hbar = h / (2 \pi)$. It is
obviously equivalent to eq.~(\ref{eq_ParaxWaveEq}) above.

Since the wave functions for free particles and those subjected to
harmonic potentials factorize with respect to their spatial
coordinates we will only discuss the one-dimensional case.

It has recently been realized that a simple coordinate
transformation~\cite{Ole_AJP05} maps paraxial beams onto
two-dimensional harmonic oscillator wave functions. Clearly, the
same coordinate transformation can be used for the mapping of a free
onto a harmonically trapped particle; in a more general setting this
was noted before~\cite{Takagi_PIT90,Bluman_JPA96}. This case has
added meaning since a non-adiabatic physical transition constitutes
an experimentally implementable sequence of changing environmental
conditions. It turns out that modelling such a transition using the
standard eigenstate-projection or quantum propagator techniques is
more cumbersome. State projection typically leads to infinite sums
over eigenfunctions which have to be truncated and are difficult to
simplify, propagator techniques involve non-trivial integrals. The
results reported here may therefore not only be of fundamental but
also of technical interest.

It is noteworthy that the discontinuities of the mapping of the
time-coordinate in~eq.~(\ref{eq_t_map}) maps from a \schr equation
with a continuous to another with a discrete energy spectrum.

\IgnoreThis{\section{Maps}}

\subsection{Map from free to trapped case}

The coordinate transformations~\cite{Takagi_PIT90}
\begin{eqnarray}
x(\xi,\tau) = \frac{\xi \sqrt{b \; \omega}}{ \cos( \tau \omega ) }
% \label{eq_x_map}
%\\
\quad \mbox{and} \quad t(\tau) = b \tan( \tau \omega ) \; ,
\label{eq_t_map}
\end{eqnarray}
applied to the wave function mapping
\begin{eqnarray}
\psi(\xi,\tau) & = & \frac{\phi(x(\xi,\tau),t(\tau)) }{
f(x(\xi,\tau),t(\tau);b)} , \label{eq_func_map}
\\
\mbox{where} \quad f(x,t;b) & = & \frac{\exp \left( { \frac{i \alpha
t x^2}{4 (t^2+b^2)}} \right) }{ (1+t^2/b^2)^{1/4}} ,
\label{eq_wavefactor}
\end{eqnarray}
yield the solution~$\psi(\xi,\tau)$ for the \schr equation
\begin{eqnarray}
-{\frac {\partial^2 }{\partial  \xi^2  }} \psi \left( \xi,\tau
\right)  +i \alpha\,{\frac {\partial }{\partial \tau}}  \psi \left(
\xi,\tau \right) + \frac{ k \alpha}{2 \hbar} \xi^2  \psi \left(
\xi,\tau \right) =0   \label{eq_HOSC_SchroedingerEq}
\end{eqnarray}
of a harmonic oscillator with mass~$M$, spring constant~$ k$ and
resonance frequency~$\omega = \sqrt{k/M}$.

Although this transformation maps $t\in [-\infty, \infty ]$ onto $\tau
\in (-\pi/2,\pi/2)$, compare with the Gouy-phase of
optics~\cite{Ole_AJP05}, the periodicity arising through the use of
trigonometric functions meaningfully represents the oscillator's
motion for all times~$\tau$.

In beam optics the confocal parameter~$b$~\cite{Pampaloni_2004}
parameterizes the strength of the beam's focussing and the curvature
of its hyperbolic flow lines~\cite{Ole_AJP05}. Here, it serves as a
rescaling parameter of the transverse coordinate transformation and
thus allows us to stretch or squeeze the width of the wave packet we
want to map.

In order to preserve the wave function normalization we have to
determine the spatial coordinate stretching at the mapping time $t_0 =
\tau_0 = 0$. This yields the normalization factor~${\cal{N}}=(b\;
\omega)^{1/4}$ which the wave function has to be multiplied with. The
specific choice
\begin{eqnarray}
b_{1} = 1/\omega = \sqrt{M/k}  \label{eq_b_natural}
\end{eqnarray}
yields the natural mapping~$x=\xi$ and~${\cal{N}}=1$ which we will
use from now on.

\subsection{Inverse map (trapped to free)}\label{sseq_InvMap}

The inverse to the coordinate transformations~(\ref{eq_t_map}) are
\begin{eqnarray}
\xi & = & \frac{x}{\sqrt{ \left( 1+ \omega^2 t^2 \right) }}
\label{eq_xi_inv_map}
\\
\mbox{and} \qquad
\tau & = & \frac{1}{\omega} \arctan\left( \omega t \right) \; ,
\label{eq_tau_inv_map}
\end{eqnarray}
and go together with the wave function multiplication~$ \phi =
\psi  f $, inverting eq.~(\ref{eq_func_map}).

\subsection{Concatenated maps (trapped to trapped)}

The concatenated coordinate transformations from an initial harmonic
trapping potential with spring constant~$k$ and wave function $\psi
(\xi,\tau;k)$, via the free particle-case~$\phi$, to a final harmonic
potential with spring constant~$K$ and wave function
$\Psi(\xi,\tau;k,K)$ is given by
\begin{eqnarray}
\Xi & = &
\frac{\xi}{\sqrt{\cos^2( \Omega \tau) +\frac{k}{K}\sin^2(\Omega \tau)}}
\; , \label{eq_xi_Xi_map}
\\
T & = & \frac{1}{\omega} \arctan \left( \sqrt{\frac{k}{K}} \tan\left(
\Omega \tau \right) \right) ,
\qquad \label{eq_tau_Tau_map}
\end{eqnarray}
\begin{eqnarray}
\mbox{and} \quad \Psi(\xi,\tau;k,K)  & = & \psi(\Xi(\xi,\tau),T(\tau)) \\
\nonumber
& & \times  \frac{f(x(\xi,\tau),t(\tau);\frac{1}{\omega}) }{
f(x(\xi,\tau),t(\tau);\frac{1}{\Omega})} . \quad \label{eq_func_map_hosc_HOSC}
\end{eqnarray}

Here $\Omega=\sqrt{K/M}$, and~$\Psi(\xi,\tau;k,K)$ solves \schr
eq.~(\ref{eq_HOSC_SchroedingerEq}) with $k$ substituted by~$K$.

As expected for a map from one harmonic oscillator to another, the
inverse of the coordinate transformations~(\ref{eq_xi_Xi_map})
and~(\ref{eq_tau_Tau_map}) are given by the same functional
expressions with the quantities pertaining to one potential swapped
with those of the other ($k \leftrightarrow K$ and $\omega \leftrightarrow \Omega$).

\IgnoreThis{\section{An Example}}

\begin{figure}[h]
\centering
\includegraphics[width=0.48\textwidth,height=0.3\textwidth]{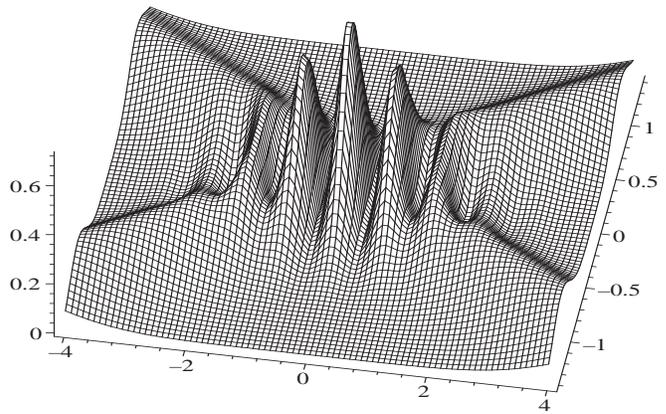}
%bb=57 648 380 720,
%\put(-260,130){\rotatebox{90}{\mbox{$ $}}}\put(-30,-10){$ $}
\caption{ Probability density $P(x,t) \sim
  |\phi_0(x,t;0,p_0)+\phi_0(x,t;0,-p_0)|^2$ of a free quantum particle
  in an equal superposition of two Gaussian states as described by
  eq.~(\ref{eq_phi_0_x_t})), with $\hbar = 1$, $M = 1$, $\sigma_0 =
  3/2$, $x_0=0$, and opposing momenta~$p_0 = 4$. The motion of the two
  half waves leads to interference at the origin.
} \label{fig_free_particle}
\end{figure}

\begin{figure}[t]
\centering
\includegraphics[width=0.48\textwidth,height=0.3\textwidth]{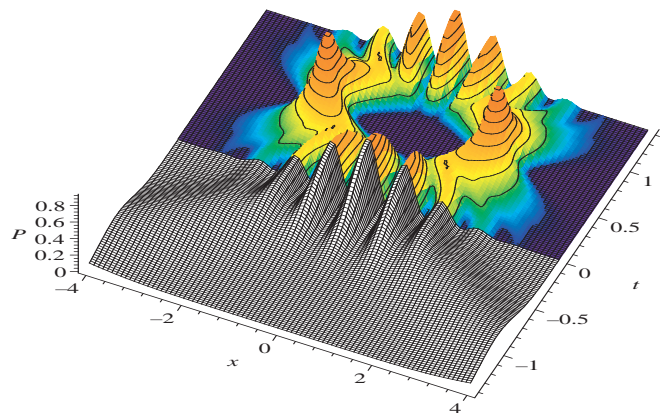}
%bb=57 648 380 720,
%\put(-260,130){\rotatebox{90}{\mbox{$ $}}}\put(-30,-10){$ $}
\caption{ Probability density $P(x,t)$ of an initially free quantum
  particle which at time~$t=\tau=0$ has the same state as the free
  particle displayed in Fig.~\ref{fig_free_particle} and suddenly gets
  trapped by a potential with spring constant $k=5$. This plot illustrates
  that the trapped particle is in a superposition state of two
  squeezed coherent states.} \label{fig_free_trapped_particle}
\end{figure}

A well-known textbook example is the freely evolving Gaussian
wave-packet with initial position spread~$\sigma_0$
\begin{eqnarray}
\phi_0(x,t;x_0,p_0)
& = & \frac {1 }{\sqrt{\sqrt{\pi} \sigma(t)}}
%\nonumber \\\exp
\exp{\left[ {{\frac {v_0\, \left( x_0 \,\hbar\,t-i M \sigma_0^{2} x
\right) }{
 \hbar\sigma_0 \sigma(t)}}}  \right]}
\nonumber \\
& \times & \exp{\left[   {-\frac { \left( x-x_0 \right) ^2}{2
\sigma_0 \sigma(t)}} -{\frac {i\sigma_0\,M{v_0}^2 t}{2 \hbar
\sigma(t)}} \right]} , \label{eq_phi_0_x_t} \\
\mbox{where} \quad \sigma(t) & = & \sigma_0+i \frac{t\hbar}{\sigma_0
M} . \label{eq_sigma_t}
\end{eqnarray}
Here $x_0$ parameterizes spatial, and $p_0 = M v_0 $ momentum
displacement of the wave functions. If either of these two quantities
is non-zero the mapping onto a harmonically trapped state results in a
state with oscillating center-of-mass. In general, although a wave
function of Gaussian shape, this freely evolving wave packet will also
not `fit' the width of the harmonic potential and therefore be
squeezed. In short, the state of eq.~(\ref{eq_phi_0_x_t}) trapped in a
harmonic potential becomes a squeezed coherent
state~\cite{Schleich_Book}. For a superposition of such states see
Fig.~\ref{fig_free_particle} and for this superposition state being
trapped consult Fig.~\ref{fig_free_trapped_particle}.

\IgnoreThis{\section{Conclusion}}

It is shown that general solutions of the free-particle \schr
equation can be mapped onto solutions of the \schr equation for the
harmonic oscillator using a simple coordinate transformation in
conjunction with a multiplication of the wave function by a suitable
phase factor. This map is invertible and a concatenation of two such
maps allows us to map from one harmonic oscillator to another with a
different spring constant. The simplicity of the approach described
here makes it a tool of choice for the description of the wave
function of a particle experiencing instantaneous transitions from a
free to a harmonically trapped state, the sudden release from a
harmonic trap, or the sudden change of the strength of its harmonic
trapping potential.

The mapping introduced here is computationally more efficient than
state-projection or propagator techniques and conceptually simpler
than mapping techniques such as those used for supersymmetric
potentials~\cite{Gendenshtein_JETP83,Cooper_JPA89}. It may even help
with numerical calculations because it allows for the determination
of the behaviour of wave functions trapped in a harmonic potential
using the lower computational overheads of wave functions in free
space.

\end{document}